\documentclass[12pt]{iopart}
\usepackage[T1]{fontenc}

\expandafter\let\csname equation*\endcsname\relax

\expandafter\let\csname endequation*\endcsname\relax

\usepackage{hyperref}
\usepackage{amsmath}
\usepackage{amsfonts}
\usepackage{siunitx}
\usepackage{physics}
\usepackage{listings}
\usepackage{changepage}
\usepackage[sorting=none, style=phys]{biblatex}
\usepackage{graphicx}
\usepackage{caption}
\usepackage{subcaption}
\begin{document}
\title[Tunnelling of a composite particle in presence of a magnetic field]{Tunnelling of a composite particle in presence of a magnetic field}

\author{B. Faulend, J. Dragašević}
\address{Department of Physics, Faculty of Science, University of Zagreb, Zagreb, Croatia}
\ead{bfaulend@yahoo.com, jan.dragasevic@gmail.com}

\begin{abstract}

We present a simple model of composite particle tunnelling through a potential barrier in presence of a (pseudo)magnetic field. The exact numerical solution of the problem is provided and the applicability to real physical systems is discussed. When the magnetic field is present some new qualitative features of the transmission spectrum are observed, compared to the previously studied composite particle tunnelling with no magnetic field. Splitting of energy levels in a magnetic field leads to splitting of transmission probability resonances, which are a generic feature of composite particle tunnelling. Magnetic field also induces precession of spin on the Bloch sphere, that can be used as a Larmor clock for measuring tunnelling time.

\end{abstract}

\maketitle

\section{Introduction}

Quantum tunnelling of systems with internal structure is a process that occurs in many physical contexts. Some examples of these processes include tunnelling of molecules \cite{Saito, Goodvin2005TunnelingOA, shegelski2005, Shegelski, Shegelski2013TunnellingOA, ShegelskiResonantTO, Krassovitskiy, Krassovitskiy2014ContributionOR, KrassovitskiyPhaseAF, Vinitsky2014ModelsOQ}, fusion of loosely-bound nuclei \cite{kimura, Bertulani_2007, ahsan, shotter, Flambaum2005QuantumTO, bacca}, tunnelling of Cooper pairs \cite{Flambaum2005QuantumTO} and tunnelling of excitons \cite{Saito1995exciton, Kavka2012TunnelingAR, Guzun2013exciton, excitons}. Generally, the composite structure of a particle undergoing tunnelling can significantly influence tunnelling probability if the dimensions of the particle become comparable to dimensions of the barrier, or if there are internal energy levels with excitation energies comparable to other energy scales of the particle.

Tunnelling of two particles in a bound state was first studied theoretically by Zakhariev and Sokolov \cite{zakhariev}. Saito and Kayanuma \cite{Saito} studied tunnelling of two particles with infinite square well binding potential through a rectangular barrier numerically and showed that resonant tunnelling of composite particles occurs generically in presence of a single potential barrier. Saito and Kayanuma also studied the tunnelling of a Wannier exciton through a single barrier heterostructure \cite{Saito1995exciton} and showed similar effects as in \cite{Saito}. Their work has been expanded by investigating more general binding potentials and barriers in \cite{penkov2000PR, Penkov2000QuantumTO, Flambaum2005QuantumTO, Goodvin2005TunnelingOA, shegelski2005, bacca}. Bertulani et al. \cite{Bertulani_2007} showed that tunnelling probability can be enhanced for systems with spin-like internal structure which can transition between states of different energy. On the experimental side, there have been numerous studies of exciton tunnelling \cite{Peterson1993, haacke, Zaitsev2008RelaxationOE, Ten1996, Emiliani1997, Heimbrodt1998, Finck2011, Crochet2011, Naouari2011}. It was shown that excitons can tunnel as a whole \cite{Lawrence1994ExcitonTR, haacke}, and resonant tunnelling behaviour was seen in \cite{haacke}.

In this paper, we expand on previous work by considering tunnelling of a composite particle with two internal degrees of freedom: binding interaction between the particles and additional two-level internal degree of freedom, formally equivalent to spin-1/2. Tunnelling of systems with two or more particles with non-zero spins in magnetic field occurs in various situations, such as nuclear fusion and tunnelling of excitons. In the latter case, due to the small binding energies, the effects of magnetic field presence can be significant. Experimental studies of exciton tunnelling in presence of the magnetic field have been conducted in \cite{Heimbrodt1998, Zaitsev2008RelaxationOE, Finck2011}. Additional motivation for studying exciton tunnelling in magnetic field comes from recent discoveries in the field of 2D magnetic materials \cite{Gong2017, Mak2019, Zhang2023} that open up new possibilities in realizing magnetic heterostructures. Spin-like degree of freedom can also be used as an internal clock for direct measurement of tunnelling time, a long-standing question in quantum physics. A recent experiment \cite{Ramos2020} managed to measure tunnelling time by using hyperfine states of rubidium atom (an effective spin-1/2 degree of freedom) as a Larmor clock \cite{Buttiker1983} in a pseudo-magnetic field. The field was generated by a laser and present only inside the barrier.

Our paper is organized as follows: in section 2 we introduce our model and the method we used for calculating transmission and reflection probabilities. In section 3 we will present and discuss our results, starting with the general features of composite particle tunnelling probabilities in magnetic field, and continuing with the results for realistic values of parameters that could correspond to exciton tunnelling through heterostructure barrier. Finally, we give our concluding remarks in section 4.


\section{Model and methods}

We consider a one dimensional two-particle system tunnelling through a rectangular barrier of width $a$. Each of the particles has a mass $m$ and is point-like. One of the particles has spin $1/2$ (or effective spin-1/2 degree of freedom), and we assume that magnetic (or pseudo-magnetic) field is present in a region of width $b$ around the barrier. The positions of particles are denoted with $x_1$ and $x_2$. If we define the centre of mass coordinate as $x = (x_1 + x_2)/2$ and the relative coordinate $y = x_2 - x_1 - l + d/2$, the Hamiltonian of our model is
\begin{equation}
    H = -\frac{\hbar^2}{4m}\frac{\partial^2}{\partial x^2} - \frac{\hbar^2}{m}\frac{\partial^2}{\partial y^2} + U(y) + V(x-y/2) + V(x+y/2) - f(x-y/2)\sigma_x
\end{equation}
Here, the interaction between particles is given as an infinite potential well:
\begin{equation}
    U(|x_2-x_1|) =
\left\{
	\begin{array}{ll}
		0,  & l-d/2 \leq |x_2-x_1| \leq l+d/2 \\
		\infty,& \mbox{otherwise},
	\end{array}
\right.
\end{equation}
where $l$ is the mean separation of the particles and $d$ is the width of the well. Potential of the rectangular barrier is defined as $V(x)=V_0 \theta(a/2-|x|)$, strength of the magnetic field is characterized by splitting of energy levels $f(x)=u\theta(b/2-|x|)$, and $\sigma_x$ denotes the Pauli-X matrix.

Our model is identical as the one in Saito and Kayanuma \cite{Saito} with the addition of interaction with the magnetic field. The qualitative features of results obtained for one dimensional composite particle model are expected to be retained in a three dimensional model, as shown in \cite{shegelski2005}, at least for the s-wave component. Magnetic field strength is approximated by a step-function which is reasonable if the field is confined to a narrow layer around the barrier. A model where only one of the particles has spin-1/2 allows us to study the effects of a general two-level internal degree of freedom on tunnelling without considering a range of other system-dependent effects. It can be justified e.g. if one of the particles has a much larger magnetic moment than the other, as is frequently the case with excitons.

Tunnelling of a composite particle through a single square potential barrier is equivalent to successive tunnelling through two potential barriers \cite{Saito}. The potential is equal to $V_0$ in the regions where $2x-a-l+d/2 \leq y \leq 2x+a-l+d/2$ and $-2x-a-l+d/2 \leq y \leq -2x+a-l+d/2$. Where these regions coalesce (which happens if $l-d/2 < a$) the potential is $2V_0$. The potential defined with these properties is denoted $W(x,y)$. Time-independent Schrödinger equation for the system is thus:

\begin{align}
\begin{split}
    -\frac{\hbar^2}{4m}\frac{\partial^2}{\partial x^2}\Psi(x,y) - \frac{\hbar^2}{m}\frac{\partial^2}{\partial y^2}\Psi(x,y) + U(y)\Psi(x,y) + \\ + W(x,y)\Psi(x,y) - f(x-y/2)
    \begin{pmatrix} 
	0 & 1\\
	1 & 0
	\end{pmatrix}
	\Psi(x,y) = E\Psi(x,y)
\end{split}
\end{align}

Wave function $\Psi(x,y)$ can be expanded in the basis of $\sigma_z$ and eigenfunctions of the infinite potential well:

\begin{equation}
    \Psi(x,y) = 
    \begin{pmatrix} 
	{\sum_{j = 1}^\infty \psi_{j1}(x)\phi_{j1}(y)}\\
	{\sum_{j = 1}^\infty \psi_{j2}(x)\phi_{j2}(y)}
	\end{pmatrix}
	=
	\sum_{j=1}^\infty
	\begin{pmatrix}
	{\psi_{j1}(x)}\\
	{\psi_{j2}(x)}
	\end{pmatrix}
	\phi_{j}(y),
\end{equation}
where $\phi_j(y) = \sqrt{\frac{2}{d}} \sin \left(\frac{j\pi}{d} y\right)$.

After substituting the expanded wave function in the Schrödinger equation, and taking the inner product with $\langle \phi_i|$, we obtain

\begin{equation}\label{important}
    \frac{d^2}{d x^2} 
    \begin{pmatrix} 
	\psi_{i1}\\
	\psi_{i2}
	\end{pmatrix}
    +k_i^2
    \begin{pmatrix} 
	\psi_{i1}\\
	\psi_{i2}
	\end{pmatrix} 
	-\frac{4m}{\hbar^2}\sum_{j=1}^\infty
	\begin{pmatrix} 
	\psi_{j1}W_{ij}(x) - \psi_{j2}F_{ij}(x)\\
	\psi_{j2}W_{ij}(x) - \psi_{j1}F_{ij}(x)
	\end{pmatrix} =
	\begin{pmatrix} 
	0\\
	0
	\end{pmatrix},
\end{equation}

where $k_j = 2\sqrt{m(E-\varepsilon_j)/\hbar}$, $\varepsilon_j = (\hbar^2/m)(j\pi/d)^2$. Functions $W_{ij}(x)$ and $F_{ij}(x)$ are defined by:
\begin{equation}
    W_{ij} (x) = \int_{-\infty}^\infty\phi_i^*(y)W(x,y)\phi_j(y) dy
\end{equation}
\begin{equation}
    F_{ij} (x) = \int_{-\infty}^\infty\phi_i^*(y)f(x-y/2)\phi_j(y) dy
\end{equation}

For an incident wave with energy $E$, where $\varepsilon_{N} < E < \varepsilon_{N+1}$, propagating states in the region out of the potential barrier can exist for channels up to $N$ (this is true for both spin up and spin down states, channel $k$ means the $k$-th eigenstate of the internal mode). The reflection amplitude $R_{ln}$ and the transmission amplitude $T_{ln}$ for the incident wave coming in channel $l$ and going out in channel $n$ are calculated using the Method of Variable Reflection Amplitude \cite{razavy}.
For a system of equations of the form

\begin{equation}\label{gen}
    \frac{d^2}{dx^2} \psi_n(x) + k_n^2 \psi_n(x) - \sum_{m=0}^\infty v_{nm}(x)\psi_m(x) = 0
\end{equation}
reflection and transmission amplitudes are solutions of the following system of differential equations:

\begin{equation}\label{R}
    \frac{d}{dx}R_{ln}(x) = -\sum_{j=0}^\infty \frac{1}{2ik_j}\left(e^{ik_jx}\delta_{lj} + R_{lj}(x)e^{-ik_j x} \right) \sum_{m=0}^\infty v_{jm}(x)\left(e^{ik_m x} \delta_{mn} + R_{mn}(x)e^{-ik_m x} \right)
\end{equation}

\begin{equation}\label{T}
    \frac{d}{dx}T_{ln}(x) = -\sum_{j=0}^\infty \frac{1}{2ik_j}T_{lj}(x)e^{ik_j x} \sum_{m=0}^\infty v_{jm}(x)\left(e^{ik_m x} \delta_{mn} + R_{mn}(x)e^{-ik_m x} \right)
\end{equation}

with boundary conditions $R_{ln}(x \rightarrow \infty) \rightarrow 0$, $R_{ln}(x \rightarrow -\infty) \rightarrow R_{ln}$, $T_{ln}(x \rightarrow \infty) \rightarrow \delta_{ln}$, $T_{ln}(x \rightarrow -\infty) \rightarrow T_{ln}$.

By solving these equations numerically for permitted channels (the sums will go from $0$ to $N$ instead of $\infty$) and in the specified range for the coordinate $x$ (the centre of mass coordinate in range $[-(2b+2l+d)/4, (2b+2l+d)/4]$), the probabilities of reflection and transmission from channel $l$ to channel $n$ can be obtained as:

\begin{align}
    P_{r,l \rightarrow n} = \frac{k_n}{k_l} |R_{ln}(-\infty)|^2 \\
    P_{t,l \rightarrow n} = \frac{k_n}{k_l} |T_{ln}(-\infty)|^2
\end{align}

\section{Results and discussion}

The most important qualitative feature of previous results \cite{Saito, Saito1995exciton, penkov2000PR, Penkov2000QuantumTO, Flambaum2005QuantumTO, Goodvin2005TunnelingOA, shegelski2005, bacca} on composite particle tunnelling is the presence of resonances, even for a single barrier. As explained in \cite{Saito}, the resonances are present because the center of mass coordinate effectively tunnels through double barrier potential $W_{ij}(x)$. When the magnetic field is present, resonant structure of tunnelling probabilities is still present, but we can also observe two additional effects: splitting of the resonance peaks and Larmor precession of spin on the Bloch sphere.

\begin{figure}[!htb]
\centering
\begin{subfigure}{.45\textwidth}
  \includegraphics[width=\linewidth]{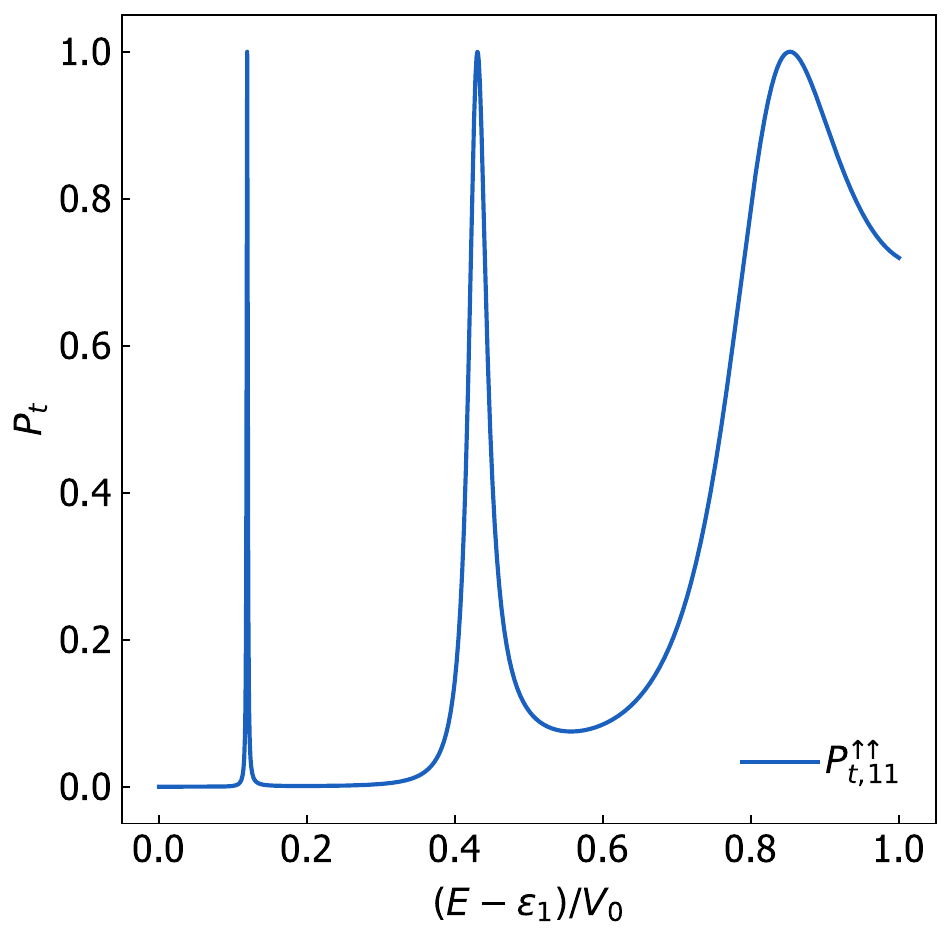}
  \caption{}
  \label{Ta}
\end{subfigure}
\begin{subfigure}{.45\textwidth}
  \includegraphics[width=\linewidth]{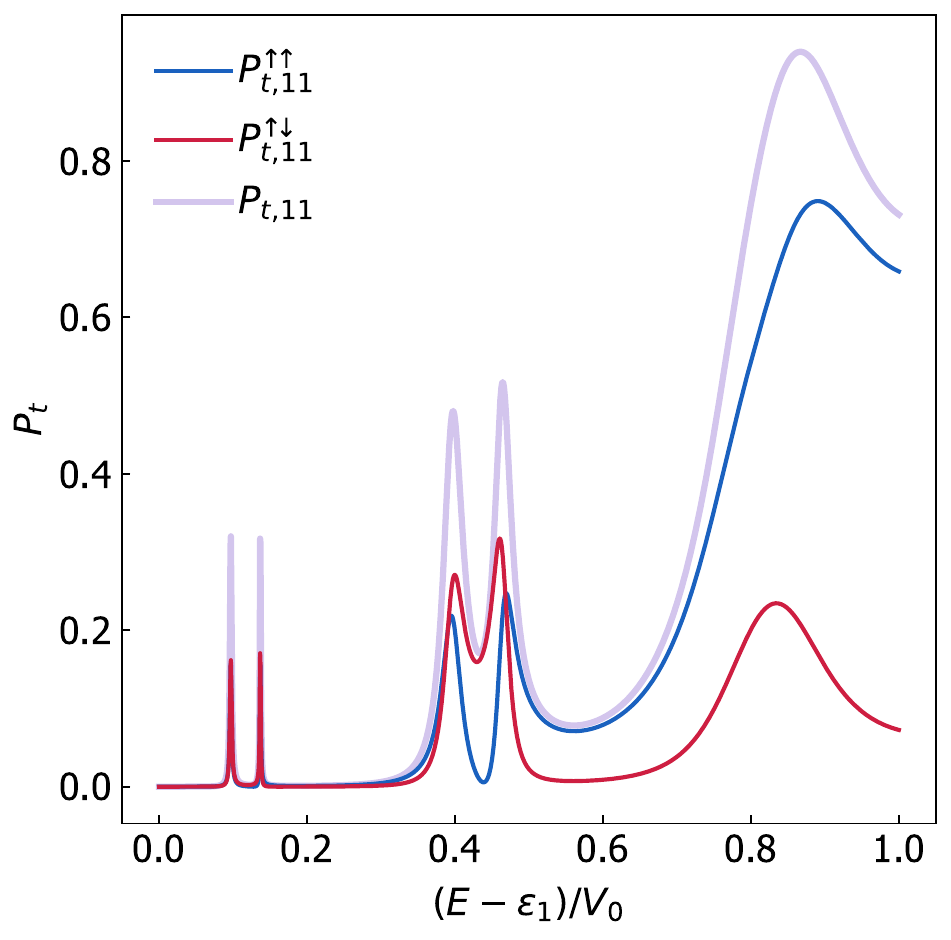}
  \caption{}
  \label{Tb}
\end{subfigure}
\caption{The graphs show transmission probabilities from the first channel for $a=1$, $b=1$, $d=5$, $l=5$ and (a) $u=0$, (b) $u=0.15$. $P_{t,11}^{\uparrow\uparrow}$ denotes a particle incoming in spin up state from channel $1$ and outgoing in the same state, $P_{t,11}^{\uparrow\downarrow}$ denotes a particle incoming in spin up state from channel $1$ and outgoing in spin down state. The probabilities $P_{t,11}^{\downarrow\downarrow}$ and $P_{t,11}^{\downarrow\uparrow}$ are the same as $P_{t,11}^{\uparrow\uparrow}$ and $P_{t,11}^{\uparrow\downarrow}$, respectively.}
\label{T11}
\end{figure}

\subsection{Splitting of resonances}

\begin{figure}[!htb]
\centering
\begin{subfigure}{.45\textwidth}
  \includegraphics[width=\linewidth]{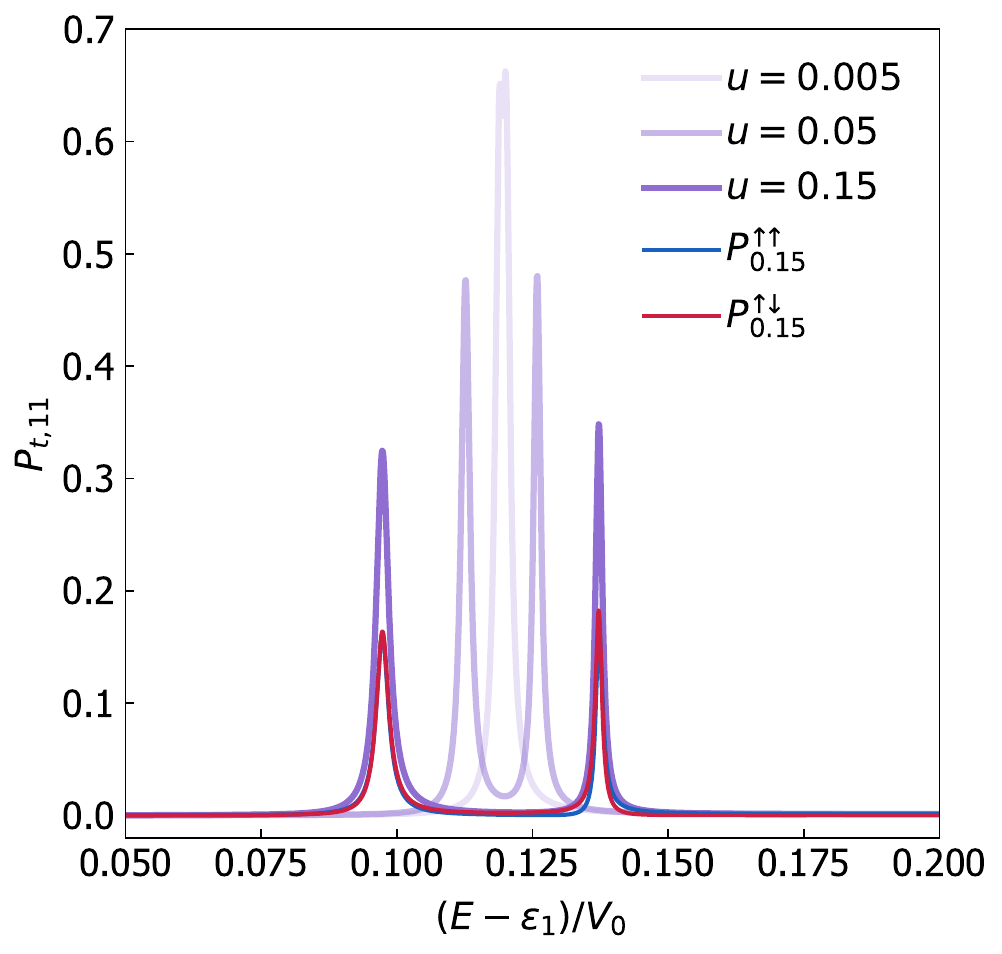}
  \caption{}
  \label{Pa}
\end{subfigure}
\begin{subfigure}{.45\textwidth}
  \includegraphics[width=\linewidth]{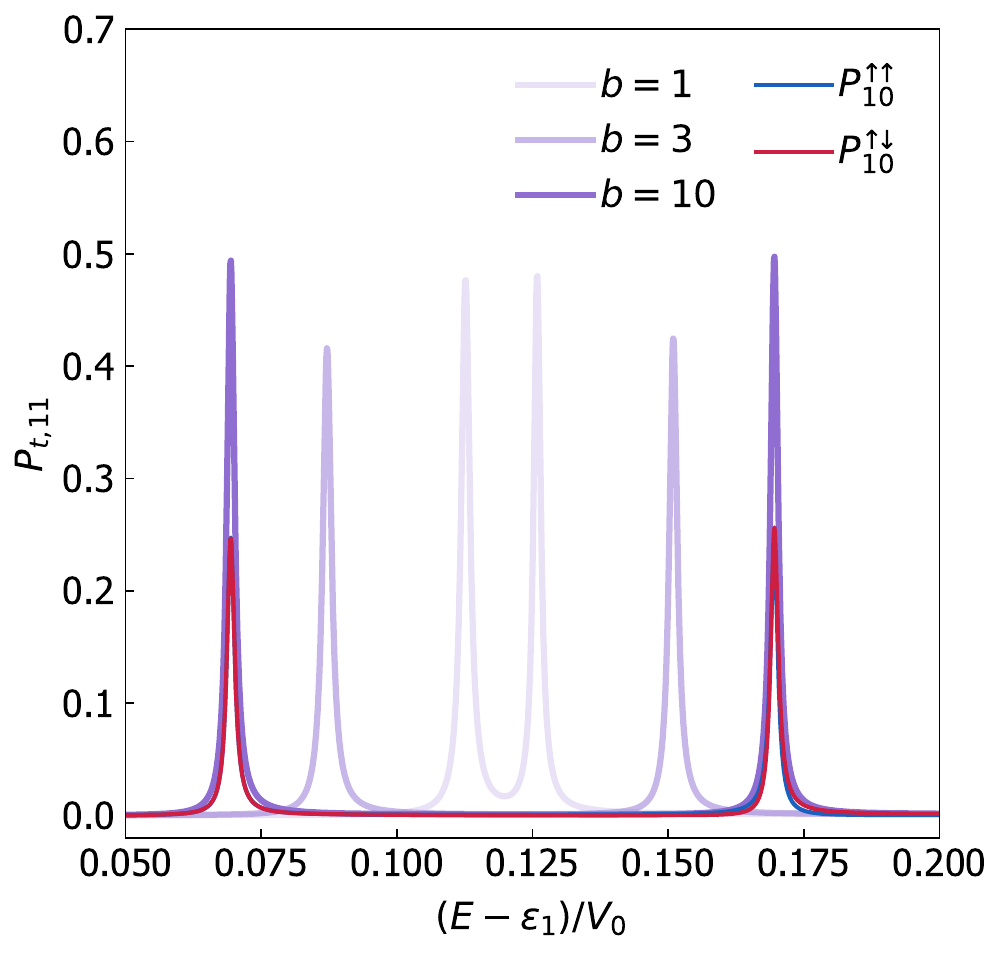}
  \caption{}
  \label{Pb}
\end{subfigure}
\caption{Total transmission probabilities close to the lowest energy resonance with the same parameters as in Figure \ref{T11}. We observe splitting of the resonances with (a) fixed magnetic field width $b=1$ and varying magnetic field strength $u=0.005$, $u=0.05$, $u=0.15$, (b) fixed $u=0.05$ and varying $b=1$, $b=3$, $b=10$. For $u=0.15$ in (a) and $b=10$ in (b), tunnelling probabilities $P^{\uparrow\uparrow}$ and $P^{\uparrow\downarrow}$ (without and with spin-flip) are also plotted.}
\label{P}
\end{figure}

A typical case of resonance splitting, which is a consequence of energy level splitting induced by the magnetic field, can be seen in Figure \ref{T11}. One of the results shown in \cite{Saito} is reproduced in Figure \ref{Ta} when the magnetic field is absent, using the variable reflection amplitude method. Keeping all other parameters the same, the effects of the magnetic field are observed in Figure \ref{Tb}.

\begin{figure}[!htb]
\centering
\begin{subfigure}{.40\textwidth}
  \includegraphics[width=\linewidth]{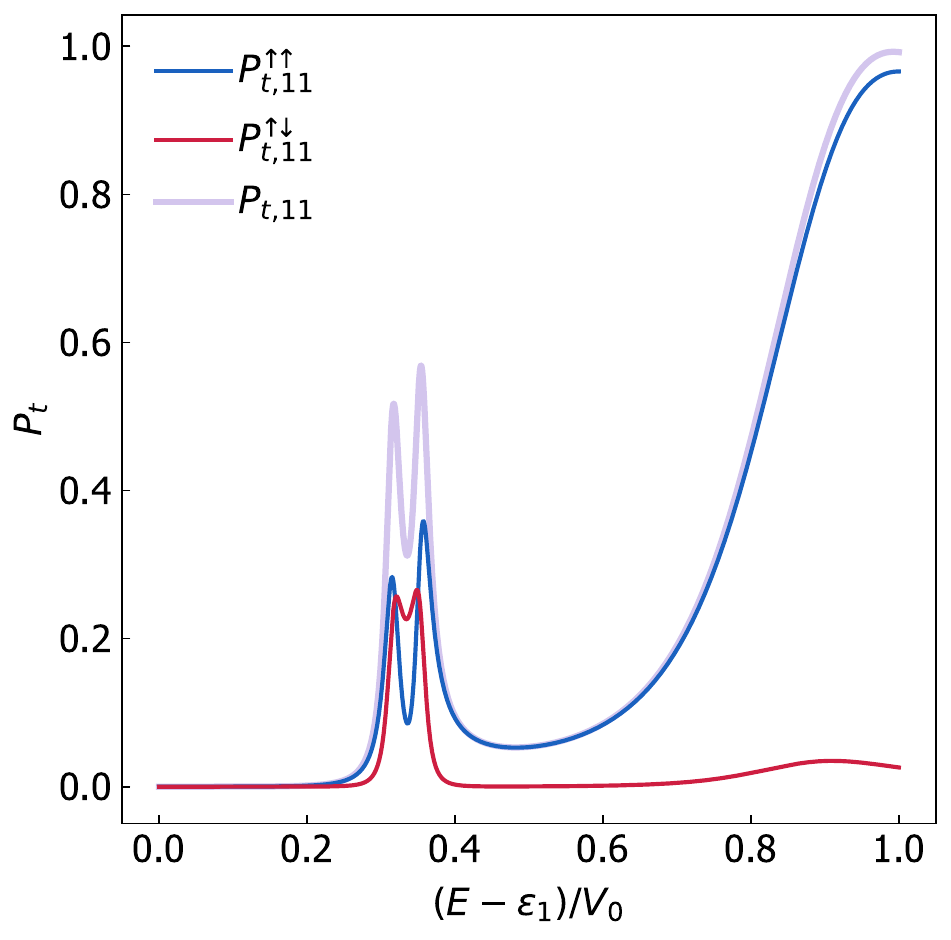}
  \caption{}
  \label{Ba}
\end{subfigure}
\begin{subfigure}{.40\textwidth}
  \includegraphics[width=\linewidth]{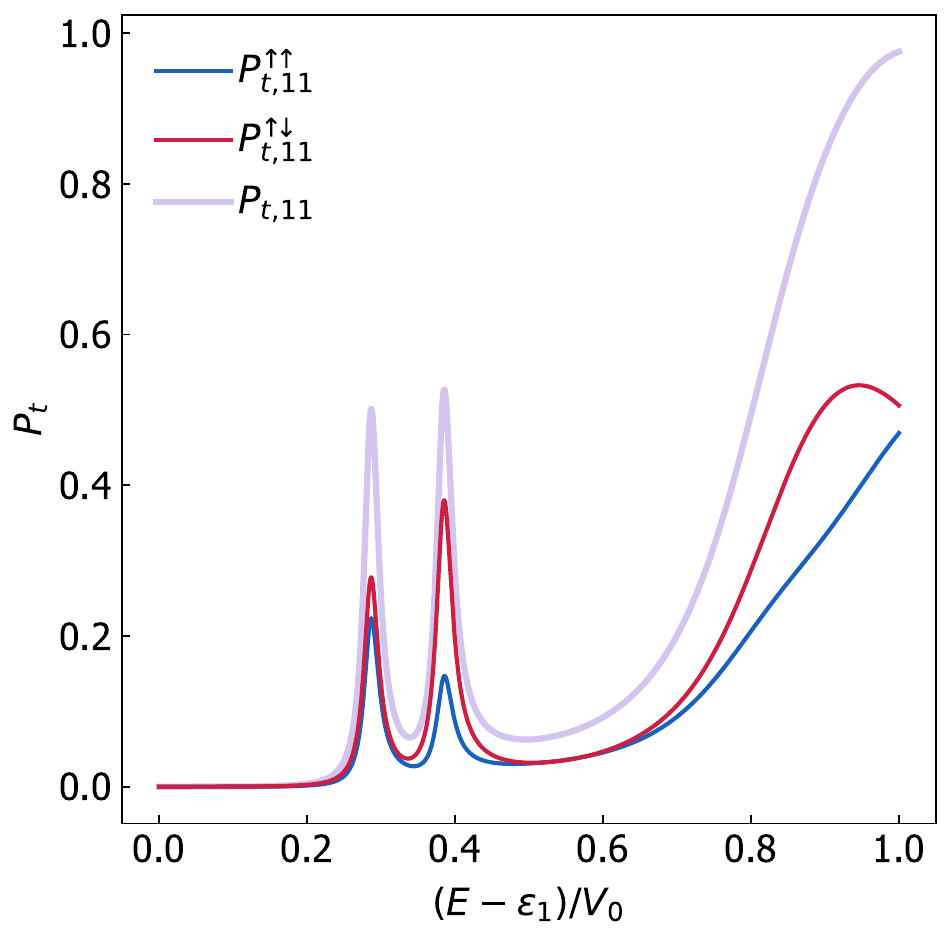}
  \caption{}
  \label{Bb}
\end{subfigure}
\begin{subfigure}{.40\textwidth}
  \includegraphics[width=\linewidth]{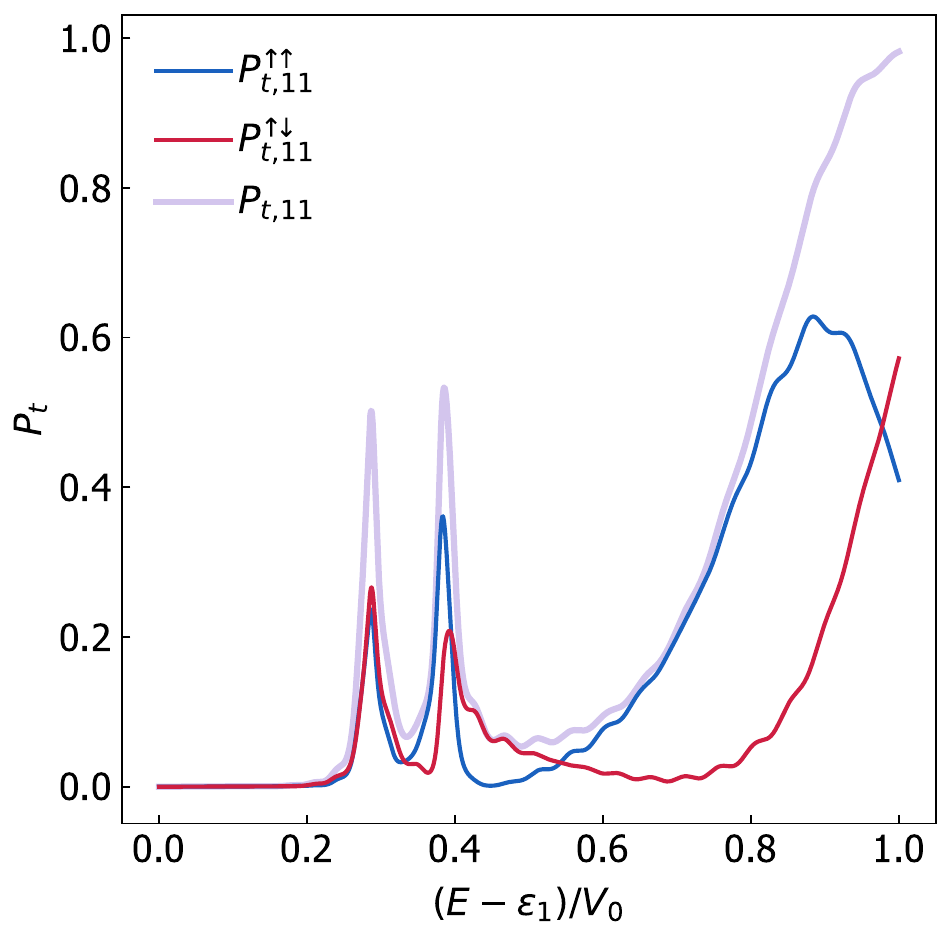}
  \caption{}
  \label{Bc}
\end{subfigure}
\begin{subfigure}{.40\textwidth}
  \includegraphics[width=\linewidth]{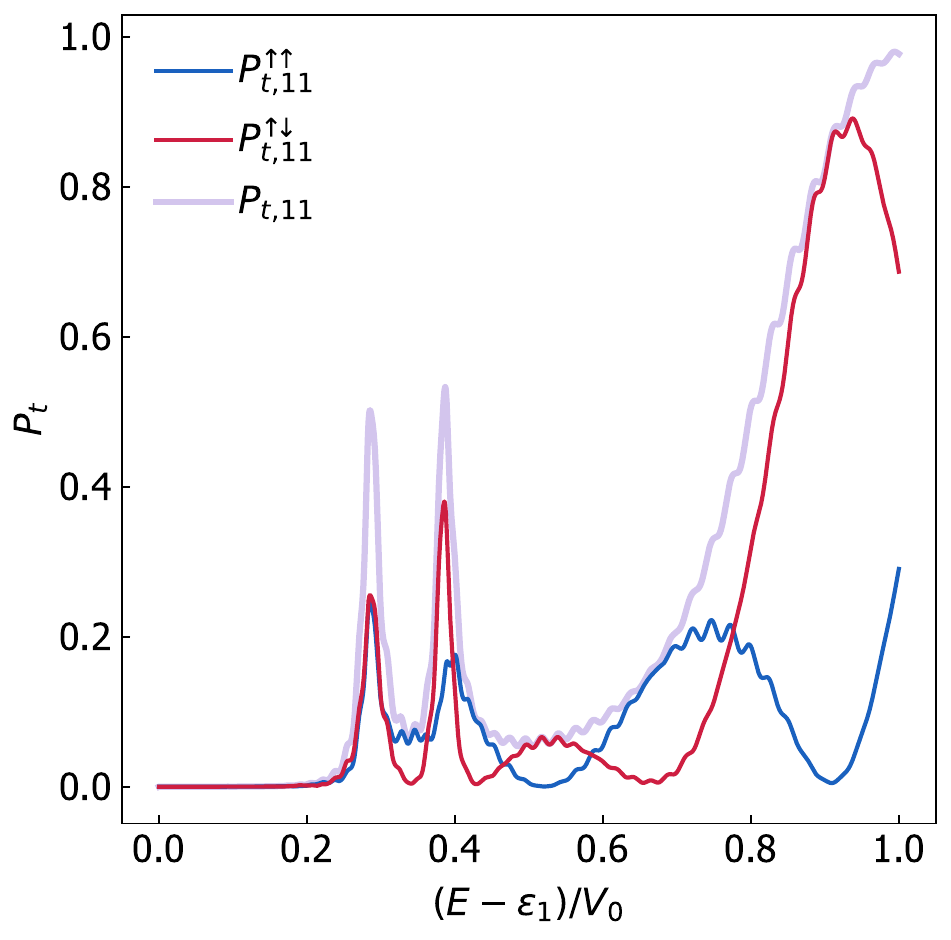}
  \caption{}
  \label{Bd}
\end{subfigure}
\caption{The graphs show transmission probabilities from the first channel for $a=1$, $d=5$, $l=3$, $u=0.05$ and (a) $b=1$, (b) $b=8$, (c) $b=100$, (d) $b=200$.}
\label{B}
\end{figure}

In the eigenbasis of $\sigma_x$, energy levels are shifted by $\pm u$ where magnetic field is present. This leads to the structure of transmission probability nearly identical to the one seen in Figure \ref{Ta}, but with the appropriate shift in energy. If we return to $\sigma_z$ eigenbasis, its eigenvectors are given by $\ket{\uparrow}=1/\sqrt{2}(\ket{+}+\ket{-})$, $\ket{\downarrow}=1/\sqrt{2}(\ket{+}-\ket{-})$, where $\ket{\pm}$ denote the eigenstates of $\sigma_x$. Thus, the tunnelling probabilities in $\sigma_z$ eigenbasis are given with $P_t^{\uparrow}=1/2(P_t^++P_t^-)$. This is most clearly seen around the lowest energy resonance where we observe 2 peaks with tunnelling probability of approximately $0.5$ (Figure \ref{Pb}). When we take the limit where the magnetic field is present everywhere (i.e. $b\to \infty$), the distance between the peaks increases and tends to $2u$ (Figure \ref{B}), which is in line with expectations. The splitting of resonance peaks can be observed for any superposition or mixture of $\ket{+}$ and $\ket{-}$ eigenstates when the field is absent. Bertulani et al. \cite{Bertulani_2007} concluded that an extremely strong magnetic field is necessary to observe significant effects on tunnelling. This is no longer the case if we examine composite particle tunnelling, as the splitting of resonance peaks can be observed for $u$ that is 2-3 orders of magnitude smaller than $V_0$ (Figure \ref{Pa}). 

\begin{figure}[!htb]
\centering
\begin{subfigure}{.45\textwidth}
  \includegraphics[width=\linewidth]{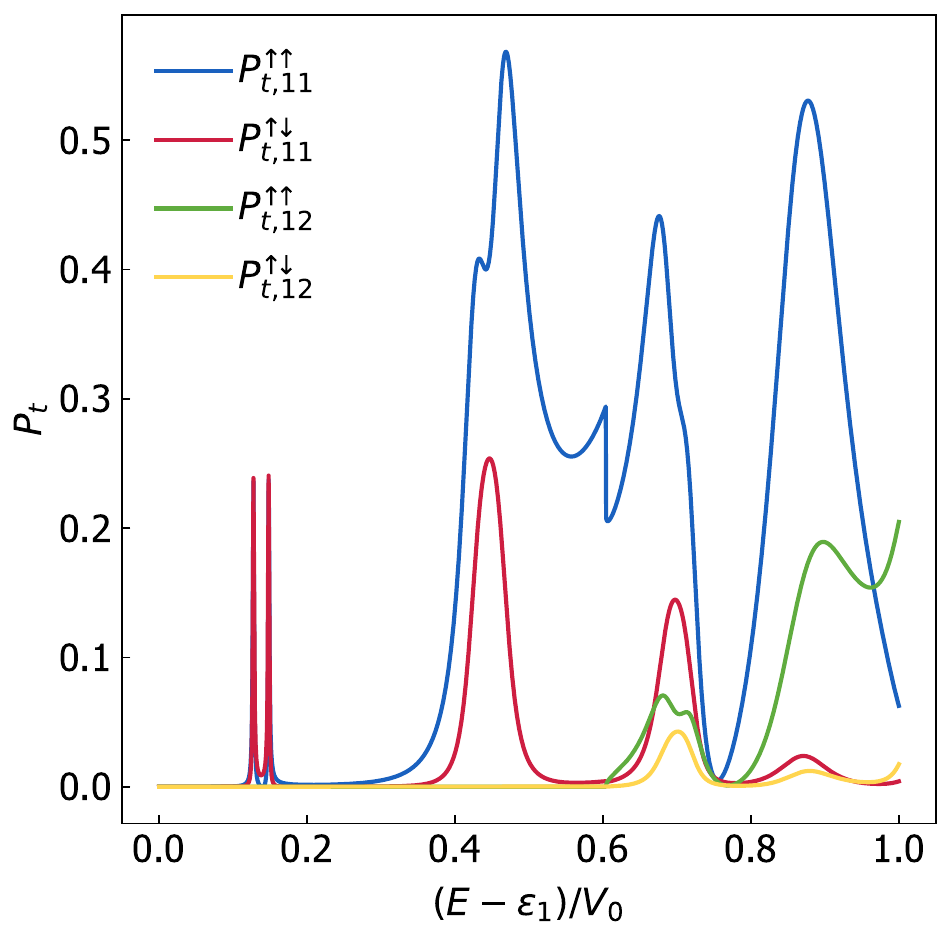}
  \caption{}
  \label{Da}
\end{subfigure}
\begin{subfigure}{.45\textwidth}
  \includegraphics[width=\linewidth]{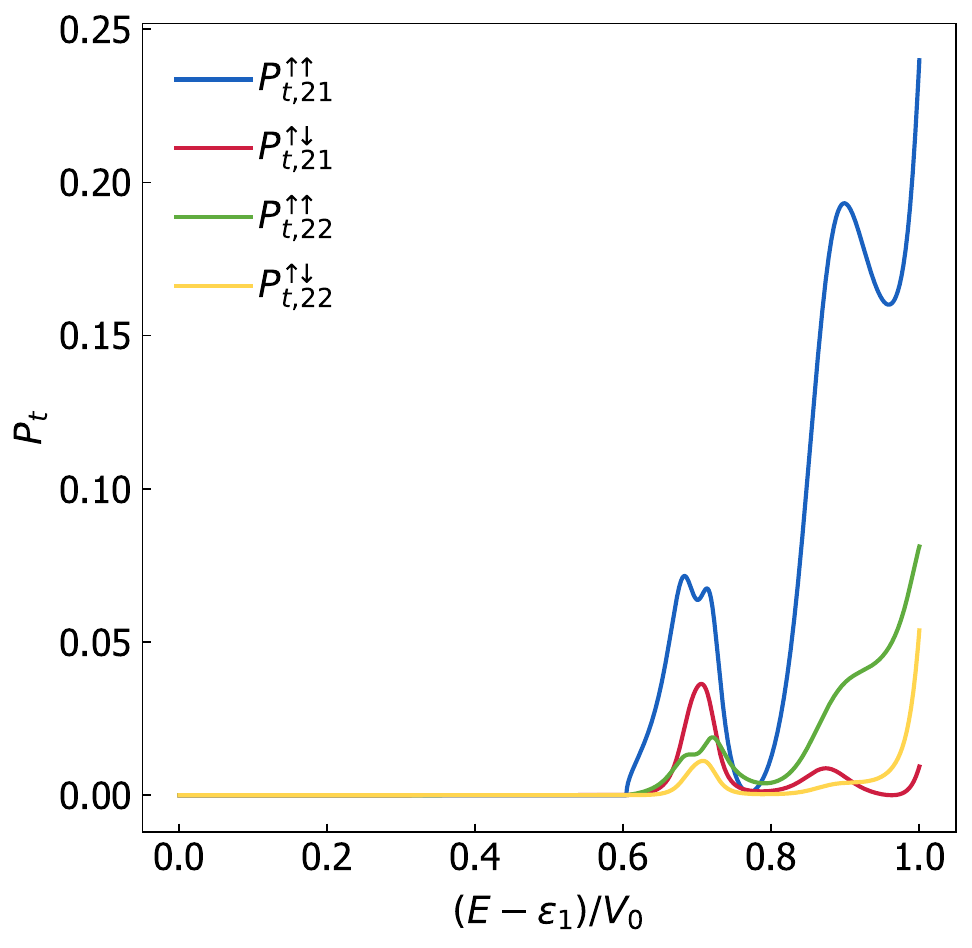}
  \caption{}
  \label{Db}
\end{subfigure}
\caption{The graphs show transmission probabilities for $a=1$, $d=7$, $l=5$ (as in \cite{Saito}), $b=1$, $u=0.05$ (a) from the first channel, (b) from the second channel. The second channel opens at the incident energy of around $0.6V_0$.}
\label{D}
\end{figure}

Results with multiple open channels can be seen in Figure \ref{D}. A discontinuity is observed in $P_{t, 11}^{\uparrow\uparrow}$ when the second channel becomes available. This is consistent with results from previous works \cite{Goodvin2005TunnelingOA}. We also observe that the total probability of transmission is reduced when higher channels open up, as in the case with no magnetic interaction \cite{Goodvin2005TunnelingOA, Bertulani_2015}.

\subsection{Larmor precession}

\begin{figure}[!htb]
\centering
\begin{subfigure}{.45\textwidth}
  \includegraphics[width=\linewidth]{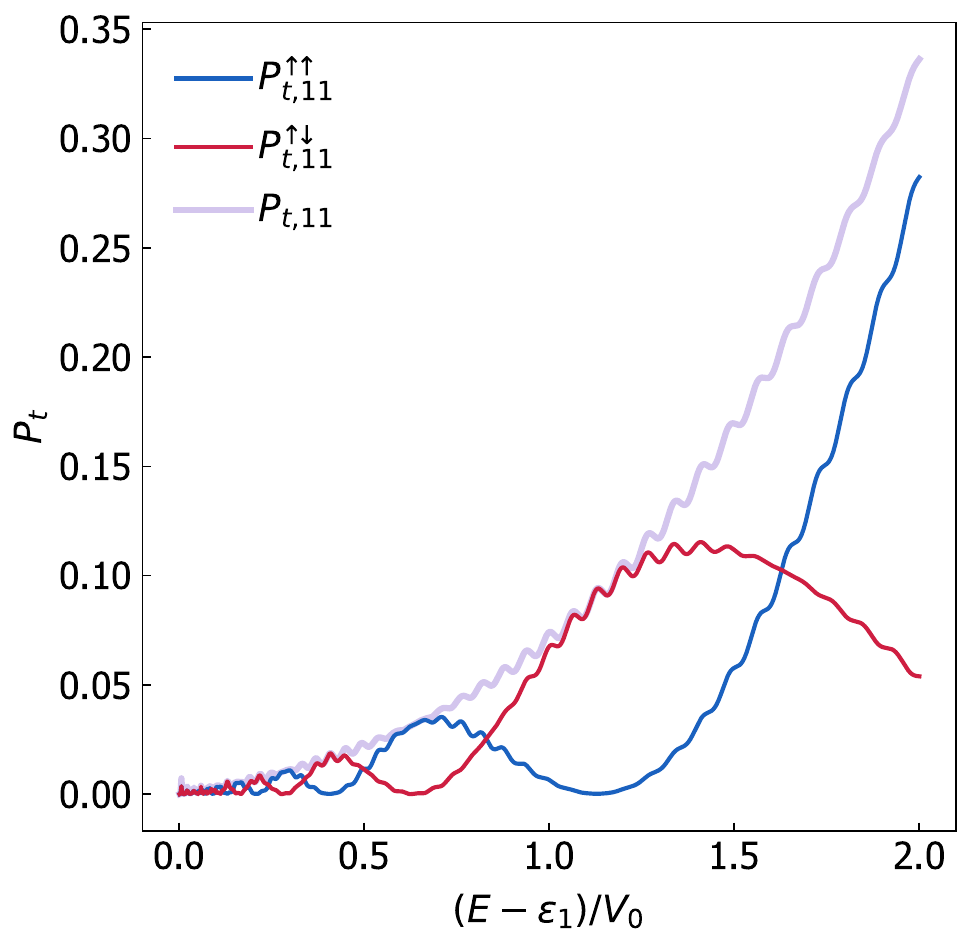}
  \caption{}
  \label{Fa}
\end{subfigure}
\begin{subfigure}{.45\textwidth}
  \includegraphics[width=\linewidth]{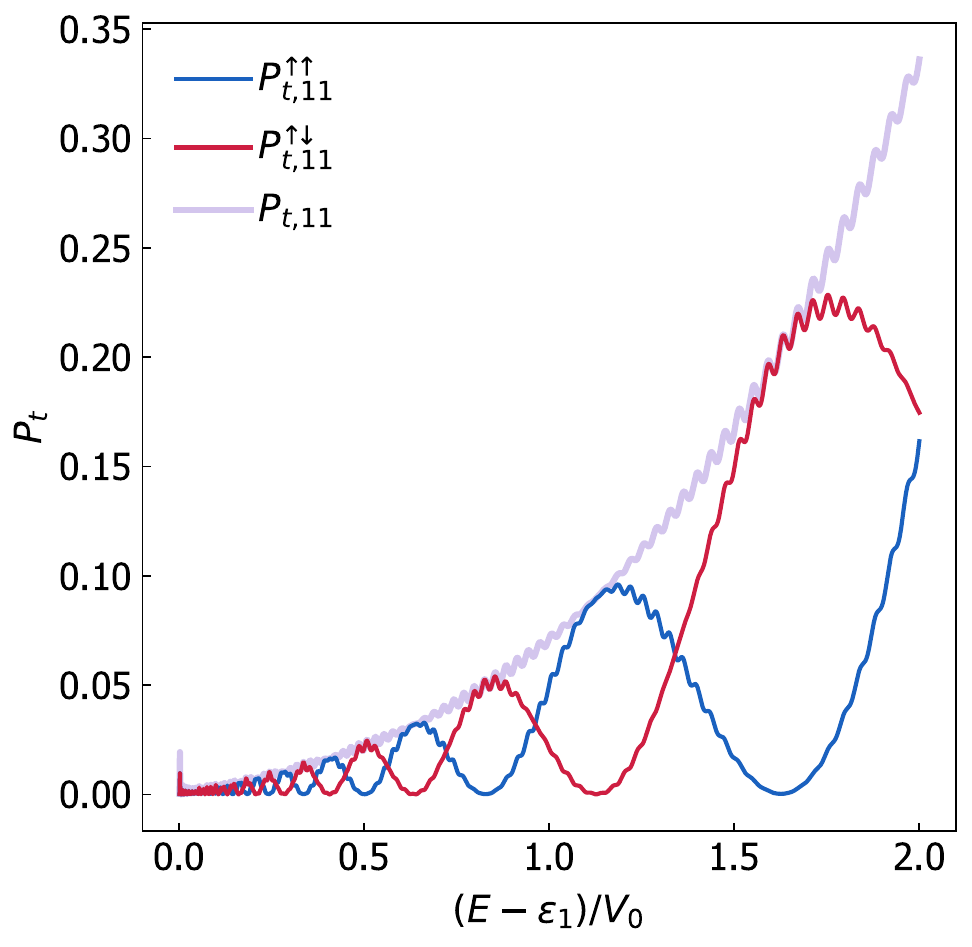}
  \caption{}
  \label{Fb}
\end{subfigure}
\caption{The graphs show transmission probabilities from the first channel for $a=1$, $u=0.05$, $d =0.05$, $l=0.05$ (a) $b=100$, (b) $b=200$.}
\label{F}
\end{figure}

Figures \ref{B} and \ref{F} show the effect of increasing the width $b$ of area where the magnetic field is present for composite and single particle tunnelling, respectively. Larger $b$ leads to oscillations of tunnelling probabilities $P_{t, 11}^{\uparrow\downarrow}$ and $P_{t, 11}^{\uparrow\uparrow}$ (tunnelling with and without spin-flip). For small $b$, $P_{t, 11}^{\uparrow\downarrow}$ is also small, except near the resonances and for small energies. These observations are a consequence of Larmor precession of spin about the direction of the magnetic field. Total angle of spin rotation on the Bloch sphere is given with $\theta=\omega_L\tau$, where $\omega_L$ is Larmor frequency and $\tau$ is the time spent inside the magnetic field. Time spent inside the magnetic field depends on $b$ and the incident energy. If $\theta=2n\pi$ with $n\in \mathbb{N}$, $P_{t, 11}^{\uparrow\downarrow}=0$, and if $\theta=(2n+1)\pi$, $P_{t, 11}^{\uparrow\uparrow}=0$. When $b$ is doubled, we expect that $\theta$ is also roughly doubled, and that the period in incident energy of spin-flip probability oscillations is halved. This expectation is confirmed by our results, as seen in Figures \ref{B} and \ref{F}.

For very large $b$, we also observe smaller oscillations in tunnelling probability, but these oscillations are not a physically meaningful effect. They are caused by our choice of a hard cutoff for the magnetic field, $f(x)\propto \theta(b/2-|x|)$, which is convenient as functions $F_{ij}(x)$ can be analytically evaluated, but a physical magnetic (or pseudo-magnetic) field will always be a smooth function of spatial coordinates. Evaluating the tunnelling probabilities with a soft cutoff for the magnetic field leads to disappearance of the aforementioned oscillations.

Close to the resonances, there is a much larger probability of tunnelling with spin-flip (Figure \ref{Pa} and \ref{B}). For example, the lowest energy resonances approximately have $P_{t}^{\uparrow\downarrow}\approx P_{t}^{\uparrow\uparrow}$. It is possible to understand this from the $\sigma_x$ eigenbasis and the relations $\ket{\uparrow}=1/\sqrt{2}(\ket{+}+\ket{-})$, $\ket{\downarrow}=1/\sqrt{2}(\ket{+}-\ket{-})$. Each resonance for $\ket{\uparrow}$ and $\ket{\downarrow}$ state corresponds to a resonance of $\sigma_x$ eigenstate $\ket{\pm}$ (there is a single resonance for $\ket{+}$ and $\ket{-}$ that reaches 100\% tunnelling probability). As $\sigma_x$ eigenstates are given with $\ket{+}=1/\sqrt{2}(\ket{\uparrow}+\ket{\downarrow})$, $\ket{-}=1/\sqrt{2}(\ket{\uparrow}-\ket{\downarrow})$, a resonance corresponding to either of $\sigma_x$ eigenstates will have 50\% spin-flip probability after tunnelling. We can also connect this result with our previous discussion of Larmor precession and tunnelling time. Tunnelling resonances occur when center of mass coordinate forms a quasi-bound state in its effective double barrier potential \cite{Saito, Bertulani_2015}. Thus, we expect that tunnelling time should be much longer than outside of the resonances, and this is confirmed in our results: significantly larger spin-flip probability directly points to a larger angle of spin rotation and longer tunnelling time.

\subsection{Exciton tunnelling}

\begin{figure}[!htb]
\centering
\begin{subfigure}{.45\textwidth}
  \includegraphics[width=\linewidth]{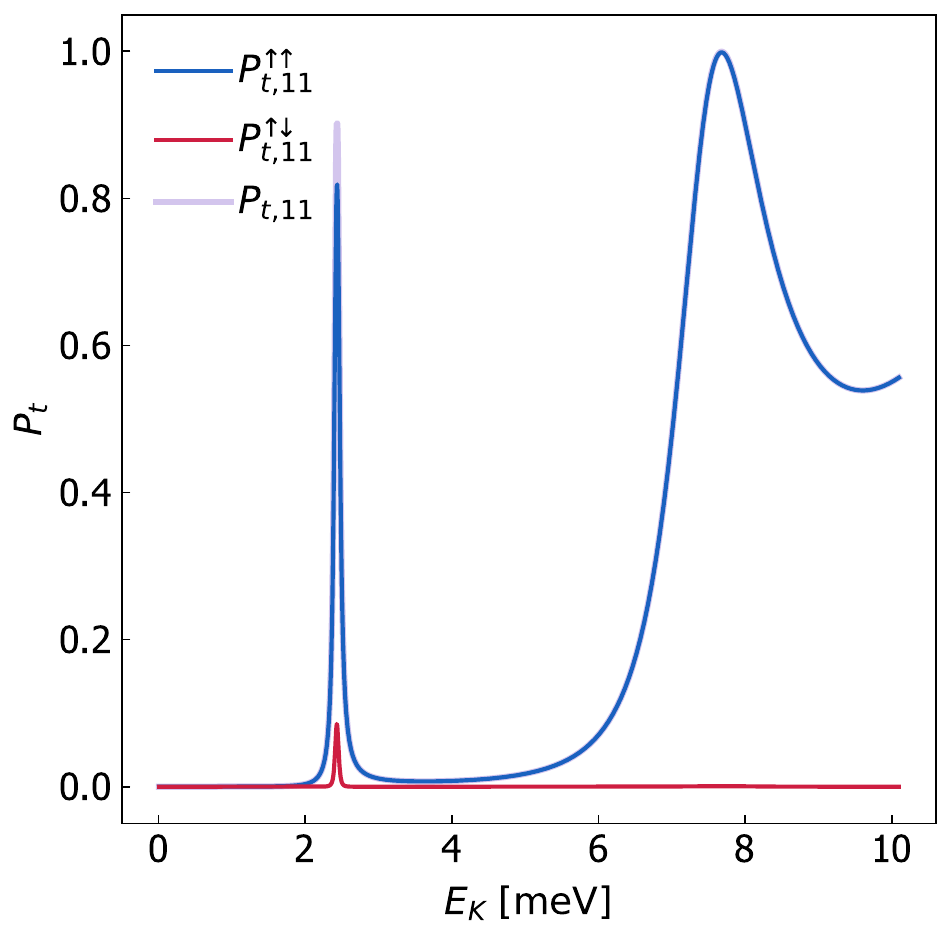}
  \caption{}
  \label{exc_a}
\end{subfigure}
\begin{subfigure}{.45\textwidth}
  \includegraphics[width=\linewidth]{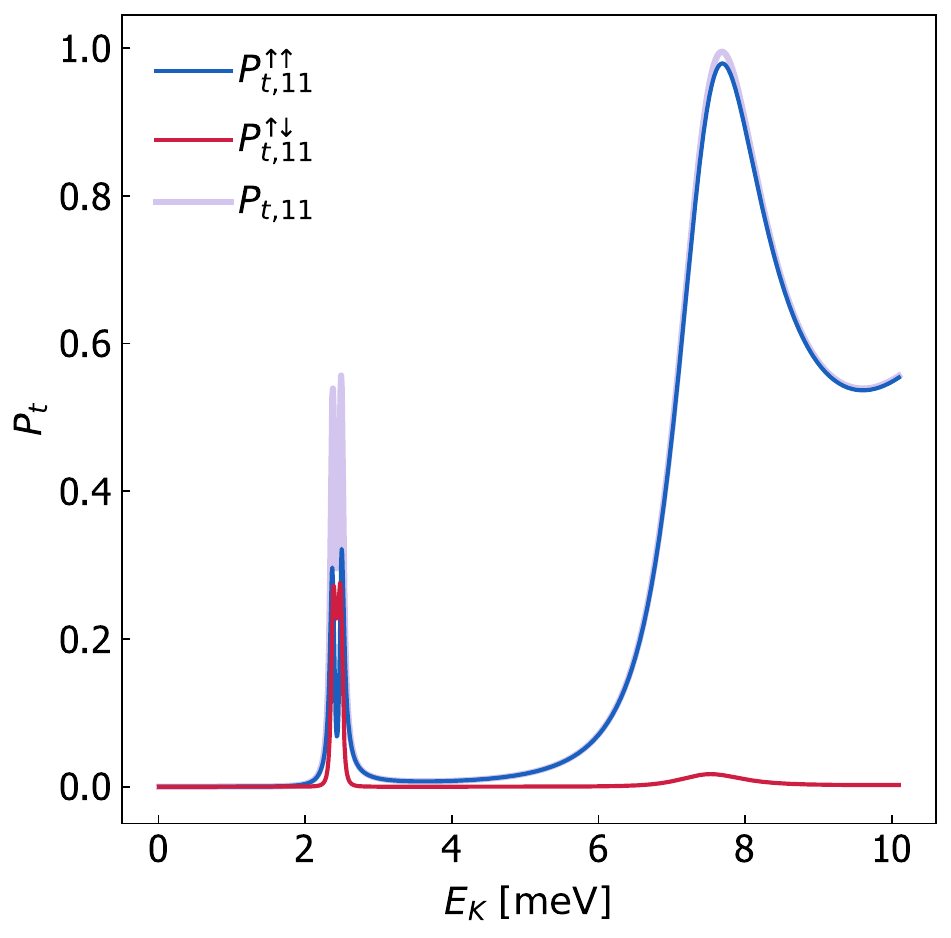}
  \caption{}
  \label{exc_b}
\end{subfigure}
\begin{subfigure}{.45\textwidth}
  \includegraphics[width=\linewidth]{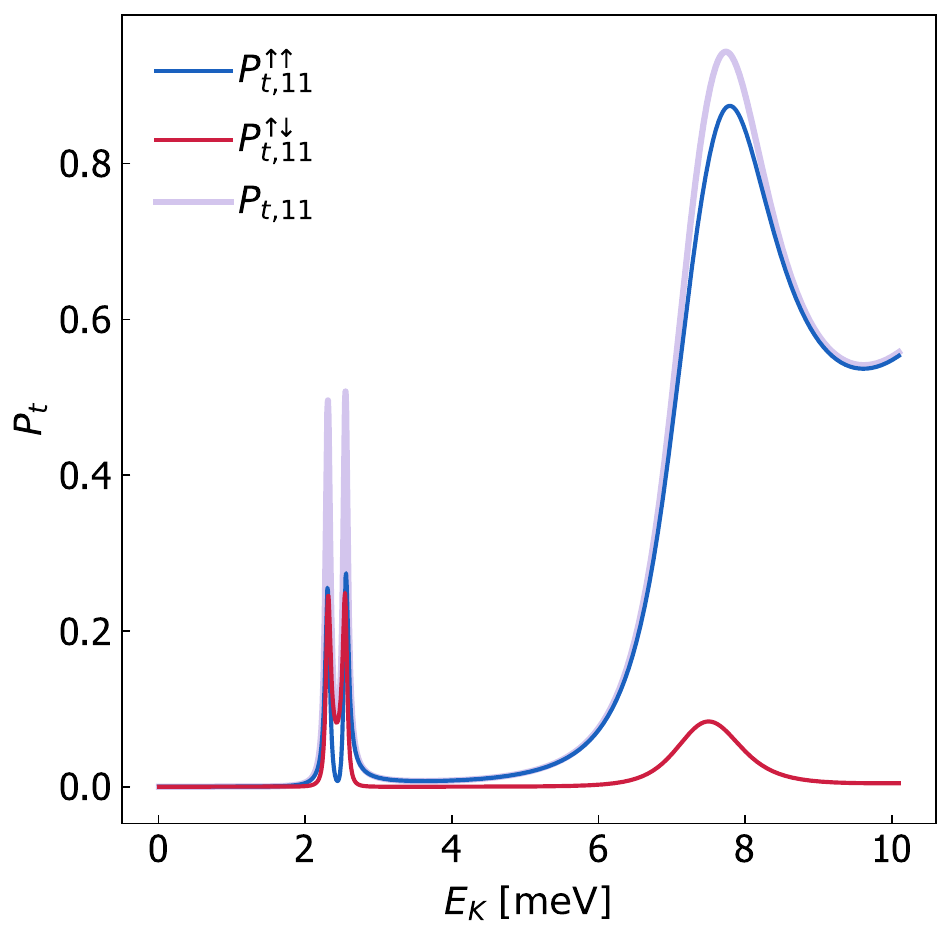}
  \caption{}
  \label{exc_c}
\end{subfigure}
\begin{subfigure}{.45\textwidth}
  \includegraphics[width=\linewidth]{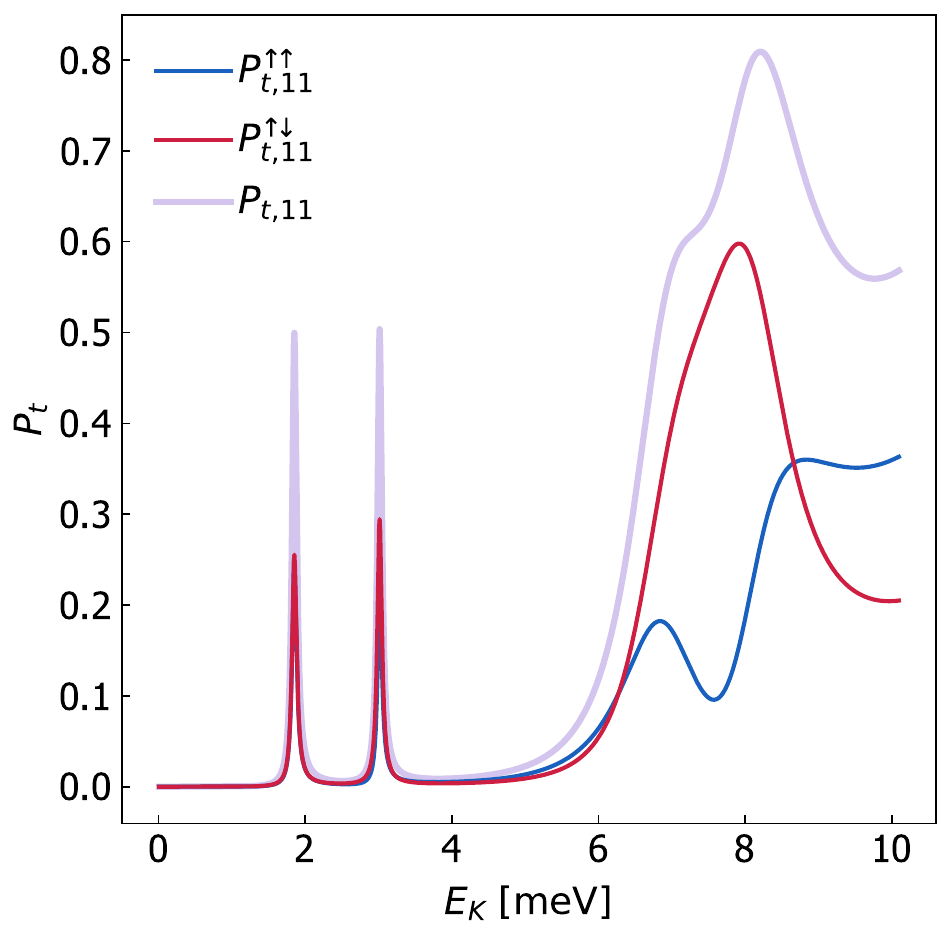}
  \caption{}
  \label{exc_d}
\end{subfigure}
\caption{Transmission probabilities for excitons, with $a=\SI{2}{nm}$, $l=\SI{10}{nm}$, $d=\SI{15}{nm}$, $m=m_e$, $V_0=\SI{15}{meV}$, $B=\SI{1}{T}$ and (a) $g_{exc}=1$, $b=a$ (b) $g_{exc}=1$, $b=8a$, (c) $g_{exc}=10$, $b=a$, (d) $g_{exc}=10$, $b=8a$.}
\label{exc}
\end{figure}

In order to investigate the effects of magnetic field on a realistic system that can be experimentally studied, we calculate tunnelling probabilities for values of parameters that are typical for tunnelling of Wannier-Mott excitons through quantum heterostructure. More specifically, we take the barrier width to be $a=\SI{2}{nm}$, and barrier height to be $V_0=\SI{15}{meV}$. Exciton Bohr radius is represented by mean distance between quasiparticles $l=\SI{10}{nm}$. We take both effective electron and hole masses to be equal to the free electron mass $m_e$, and exciton binding energy to be $\SI{10}{meV}$. In order to obtain the potential well width $d$, we approximated the binding energy with energy difference of lowest two levels $\varepsilon_2-\varepsilon_1=3\pi^2\hbar^2/md^2$, which gives $d=\SI{15}{nm}$. Splitting of energy levels in the magnetic field is given with $u=g_{exc}\mu_B B$.

The results we obtained are shown in Figure \ref{exc}. We used $g$ values $g_{exc}=1$ and $g_{exc}=10$, and width of magnetic field $b=a$ and $b=8a$. Qualitatively, results show the same effects that were discussed earlier, and splitting of resonance peaks can be clearly seen for $g_{exc}=10$ with magnetic field $B=\SI{1}{T}$. In realistic systems with giant Zeeman effect such as $\mathrm{Ga}_{1-x}\mathrm{Mn}_x\mathrm{N}$, $g_{exc}$ can be even several times higher \cite{Pacuski2007}, so we conclude that the magnetic field should have observable effects on exciton tunnelling. The effects of Larmor precession are also visible (Figure \ref{exc_d}), but experimental control of exciton spin direction is probably beyond current experimental reach.

Naturally, it would be more realistic to calculate the tunnelling probabilities with a $1/r$ interaction between the electron and the hole. Barrier height and effective masses are also in general different for electron and hole quasiparticles. However, in line with previous research \cite{Saito, Saito1995exciton, penkov2000PR, Penkov2000QuantumTO, Flambaum2005QuantumTO, Goodvin2005TunnelingOA, shegelski2005, bacca, Kavka2012TunnelingAR}, we expect the main qualitative properties like the splitting of resonances in magnetic field to be independent of specific choice of the inter-particle potential. Also, interaction of excitons with the magnetic field can lead to a range of complex behaviours, such as exciton mass increase in the magnetic field \cite{Bodnar2017} and strong dependence of exciton magnetic moment on its kinetic energy \cite{Davies2006, Kochereshko2008}. In fact, it is not possible to treat magnetic moments of electron and hole independently \cite{Davies2006}. Again, to obtain the qualitative predictions in the framework of our model, we used a realistic value of $g_{exc}$ and magnetic field strength $B$ of $\SI{1}{T}$ which is realistic in 2D magnetic heterostructures.

\section{Conclusion}

We have studied tunnelling of a composite particle in presence of a (pseudo)magnetic field. The exact numerical solution is provided within the framework of a simple model we used. Tunnelling of a composite particle leads generically to resonances in transmission probability \cite{Saito, Saito1995exciton, penkov2000PR, Penkov2000QuantumTO, Flambaum2005QuantumTO, Goodvin2005TunnelingOA, shegelski2005, bacca} due to formation of a quasi-bound state around the barrier. We showed that the presence of additional two-level degree of freedom causes splitting of the resonances, which is a consequence of energy level splitting in a magnetic field. We also showed that the presence of a magnetic field leads to Larmor precession of the spin-like degree of freedom, that represents itself in our results through the oscillations of spin-flip probability with incident energy and width of magnetic field. In principle, Larmor precession could be used to measure the tunnelling time of composite particles (as was recently done for rubidium atoms \cite{Ramos2020}) if an appropriate experimental platform is found.

Finally, we present a calculation of transmission probabilities with values of relevant parameters that correspond to exciton tunnelling through a quantum heterostructure. Due to the discovery of 2D magnetic heterostructures \cite{Gong2017, Mak2019, Zhang2023} and a possibility of high $g$ values \cite{Pacuski2007, Davies2006, Kochereshko2008}, we expect that the effects discussed in this paper could be observed in exciton tunnelling experiments. Tunnelling of molecules is another process where effects of inter-particle interactions affect tunnelling \cite{Saito, Goodvin2005TunnelingOA, shegelski2005, Shegelski, Shegelski2013TunnellingOA, ShegelskiResonantTO, Krassovitskiy, Krassovitskiy2014ContributionOR, KrassovitskiyPhaseAF, Vinitsky2014ModelsOQ}. Effects of resonance splitting and Larmor precession could be observed if two degenerate (or nearly degenerate) molecular energy levels are used as an effective spin-1/2 and coupled in the vicinity of barrier.

As discussed in previous sections, we expect the general features of our results to be retained for more realistic binding potentials and in three-dimensional space. However, it is possible that the coupling of magnetic field with orbital angular momentum in three dimensions would lead to some new effects that are not discernible in our one-dimensional model.

\section*{Acknowledgements}
We would like to thank M. Milin and M. T. Cvitaš from the University of Zagreb and also K. Pisk from Ruđer Bošković Institute for useful discussions and comments.

\printbibliography
\end{document}